\documentclass[twocolumn,showpacs,preprintnumbers,amsmath,
  amssymb,showkeys,floatfix,sort&compress,pre,dvipdfmx]{revtex4-1}
\usepackage{graphicx}
\usepackage{bm}
\usepackage{cancel}
\usepackage{hyperref}

\newcommand{\mb}[1]{\mathbf{#1}}
\newcommand{\curl}{\nabla \times}
\newcommand{\pd}[2]{\frac{\partial #1}{\partial #2}}
\newcommand{\edge}[3]{\left(#1 \xrightarrow{#2} #3\right)}

\newcommand{\sink}[3]{\left(#1 \xrightarrow{#3} #2 \xleftarrow{#1} #3\right)}
\newcommand{\beat}[3]{\left(#1,#2 \rightarrow #3\right)}
\newcommand{\innerprod}[2]{\left\langle #1,#2 \right\rangle}
\newcommand{\responsenote}[1]{}

\begin{document}

\title{On the saturation of non-axisymmetric instabilites \\ of
  magnetized spherical Couette flow}
\author{E. J. Kaplan}
\affiliation{Helmholtz-Zentrum Dresden-Rossendorf}
\email{e.kaplan@hzdr.de}

\begin{abstract}
  We numerically investigate the saturation of the hydromagnetic
  instabilities of a magnetized spherical Couette flow. Previous
  simulations \cite{Hollerbach.RSPA.2009} demonstrated a region where
  the axisymmetric flow, calculated from a 2-D simulation, was
  linearly unstable to nonaxisymmetric perturbations. Full, nonlinear,
  3d simulations \cite{Hollerbach.RSPA.2009, Travnikov.ActaMech.2011}
  showed that the saturated state would consist only of harmonics of
  one azimuthal wave number, though there were bifurcations and
  transitions as nondimensional parameters (Re, Ha) were varied. Here,
  the energy transfer between different aziumthal modes is formulated
  as a network. This demonstrates a mechanism for the saturation of
  one mode and for the suppression of other unstable modes. A given
  mode grows by extracting energy from the axisymmetric flow, and then
  saturates as the energy transfer to its second harmonic equals this
  inflow. At the same time, this mode suppresses other unstable modes
  by facilitating an energy transfer to linearly stable modes.
\end{abstract}
\maketitle
\section{Introduction}
\label{section:Introduction}

Two spheres, one inside the other, in differential rotation with a
layer of fluid between will generate a broad array of possible
dynamics in the enclosed fluid, depending on the aspect ratio, the
rotation rates of the spheres, and the viscosity of the fluid. If the
fluid is electrically conducting and permeated by a magnetic field,
applied and/or self-excited, the array of possible dynamics broadens
further. The configuration, known as magnetized spherical Couette
flow, was first studied numerically by Hollerbach
\cite{Hollerbach.RSPA.1994} as an extension of the nonmagnetic
spherical Couette problem \cite{Stewartson.JFM.1966,
  Proudman.JFM.1956}. Since then the flow has been investigated,
numerically \cite{Gissinger.PRE.2011, Hollerbach.RSPA.2001,
  Hollerbach.RSPA.2009, Wei.PRE.2008, Liu.PRE.2008, Dormy.JFM.2002,
  Dormy.EPSL.1998} and experimentally \cite{Brito.PRE.2011,
  Kelley.PRE.2010, Nataf.PEPI.2008, Sisan.PRL.2004}, under a variety
of imposed fields and magnetic boundary conditions with sometimes
surprising results. For example, in the case of a conducting inner
boundary an applied magnetic field can induce a flow rotating faster
than the inner sphere or rotating in the opposite direction to the
inner sphere, depending on the applied field configuration
\cite{Hollerbach.RSPA.2001}. \responsenote{2.1} The superrotating case
was experimentally demonstrated in the Derviche Tourneur Sodium (DTS)
experiment \cite{Dormy.EPSL.1998}. A compendium of magnetized
spherical Couette results can be found in
\cite{Hollerbach2013.Chapter7}.

\begin{table*}
  \begin{tabular}{r||l|l|l|l}
    \hline
    & Maryland \cite{Sisan.PRL.2004} & DTS \cite{Brito.PRE.2011} & Dresden & simulations \\ \hline
    fluid & Na & Na & GaInSn &  \\ \hline
    $\nu$, viscosity $(m^2s^{-1})$ & $\rm{7.4 \times 10^{-7}}$ & $\rm{7.4 \times 10^{-7}}$ & $\rm{2.98 \times 10^{-7}}$ & \\ \hline
    $\rho$, density ($kg \; m^{-3}$) & 927 & 927 & 6360 & \\ \hline
    $\sigma$, conductivity ($Ohms^{-1} m^{-1}$) & $\rm{1.0 \times 10^7}$ & $\rm{1.0 \times 10^7}$ & $\rm{3.1 \times 10^6}$ & \\ \hline
    $r_1$, inner radius ($cm$) & 5 & 7.4 & 3 or 4.5 &\\ \hline
    $r_2$, outer radius ($cm$) & 15 & 21 & 9 & \\ \hline
    $\Omega$, inner sphere rotation $rad\; s^{-1}$ & 8 & 25 & 0.01 & \\ \hline
    $B_0$, applied Magnetic Field (mT) & $<$ 90 axial & 62 dipole & $<$ 160 axial & axial \\ \hline
    $\eta$, aspect ratio & 0.33 & 0.35 & 0.33 or 0.5 & 0.33 or 0.5 \\ \hline
    Re, Reynolds Number ($\Omega r_1^2/\nu$) & $1.3 \times 10^6$ & $10^5$ & $10^3$ & $\leq1500$ \\ \hline
    Rm, Magnetic Reynolds Number ($\mu_0 \sigma \Omega r_1^2$) & $4$ & $10$ & $10^{-3}$ & $0$, by construction \\ \hline
    Ha, Hartmann Number ($B_0 r_1 \sigma^{1/2} \rho^{-1/2} \nu^{-1/2}$) & $5 \times 10^2$ & $5 \times 10^2$ & $< 1.6 \times 10^2$ & $< 100$ \\ \hline
    \hline
  \end{tabular}
  \caption{List of typical dimensional and nondimensional parameters for the
    first Maryland experiment \cite{Sisan.PRL.2004}, DTS
    \cite{Brito.PRE.2011}, the (under construction) HZDR experiment,
    and the simulations performed here. Fluid parameters for liquid
    Sodium and GaInSn are taken from \cite{Morley.RSI.2008}. \label{table:experiments}}
\end{table*}

\begin{figure}
  \centering
  \includegraphics[width=3in]{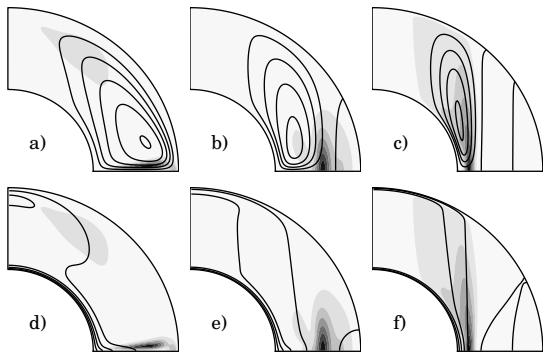}
  \caption{\label{fig:profiles} Profiles of the the energy density of
    the most unstable eigenmode from an LNSE analysis of flows at
    three different Hartmann numbers at Re 1100. a-c show streamlines
    of the meridional flow over the energy density of the m=2
    harmonic. d-f show contours of the angular momentum over the
    same. a and d show the equatorially antisymmetric jet instability
    (Re 1100, Ha 10). b and e show the equatorially symmetric return
    flow instability (Re 1100, Ha 30). c and f show the equatorially
    symmetric Shercliff layer instability (Re 1100, Ha 70).}
\end{figure}

A long, albeit contentiously, discussed result of magnetized spherical
Couette flow is the observation of an angular momentum transporting
instability in a turbulent (Re $\approx {\rm 10^7}$ ) liquid metal
flow, induced by an applied axial magnetic field, that was described
in \cite{Sisan.PRL.2004} as the long sought-after Magnetorotational
Instability (MRI). This would be momentous as the MRI is commonly
considered the mechanism by which angular momentum is removed from
accretion disks around black holes, allowing matter to fall into the
center. This is also potentially relevant to angular momentum
transport in protoplanetary disks. The instability is driven by
magnetic tension, which links together fluid parcels so that a parcel
that moves outward is azimuthally accelerated, thus being pushed
further outward, and a parcel that moves inward is azimuthally
decelerated, thus being pulled further inward
\cite{Ji.PhysicsToday.2013}. In contrast to the MRI as usually
described \cite{Balbus.RMP.1998, Balbus.ApJ.1991}, the instability
measured in \cite{Sisan.PRL.2004} was nonaxisymmetric and demonstrated
an equatorial symmetry whose parity depended on the strength of the
applied magnetic field. Subsequent numerical investigations
\cite{Gissinger.PRE.2011, Hollerbach.RSPA.2009} turned up a collection
of inductionless instabilities---related to the hydrodynamic jet
instability, the Kelvin-Helmholtz-like Shercliff layer instability and
a return flow instability---that replicated these parity transitions,
as well as the torque on the outer sphere (the proxy measurement of
angular momentum transport). Fig.~\ref{fig:profiles} shows the
streamlines of meridional circulation and isocontours of angular
momentum for the axisymmetric background flow over contours of the
energy densities of the various instabilites. These instabilities were
found by first evolving a two-dimensional (axisymmetric) flow to
steady state at a given (Re, Ha), and then applying a linearized
Navier-Stokes (LNSE) calculation to find the
fastest-growing/slowest-decaying eigenmode (in a manner similar to
\cite{Hollerbach.RSPA.2009}). A more modestly scaled (Re $< {\rm
  10^5}$), but more comprehensively diagnosed [Ultrasonic Doppler
  Velocimetry (UDV), electric potential measurements], spherical
Couette experiment is being carried out at the Helmholtz-Zentrum
Dresden-Rossendorf in order to better characterize these
instabilities, their criteria, and their saturation. Towards that end,
the Hollerbach Code \cite{Hollerbach.IJNMF.2000} is being run to
predict the signatures of the various instabilities in the available
diagnostics. Presented here is a spectral analysis of the simulations,
whose intent is to explicate the saturation and transition of
Shercliff and return flow instabilities through a comparatively small
number of nonlinear interactions. Table \ref{table:experiments} lists
dimensional and nondimensional parameters of the under construction
experiment, the simulations presented here, and two other spherical
Couette experiments for comparison.

The flow is driven by the rotating inner sphere and evolves according
to the incompressible Navier-Stokes Equation
\begin{align}
  \nabla \cdot \mb U &= 0 \nonumber \\
  \curl \mb U &= \bm{\omega} \label{eqn:NavierStokes} \\
  \pd{\bm{\omega}}{t} &= \curl \mb F + \nabla^2 \bm{\omega} \nonumber.
\end{align}
\noindent The body force $\mb F$ is given by 
\begin{equation}
  \mb F = \rm{Re} \left(\curl \mb U\right) \times \mb U
  + \rm{Ha^2} \left(\curl \mb B\right) \times \mb B \label{eqn:Forcing},
\end{equation}
\noindent with $\mb U$ and $\mb B$ vector fields of the velocity and
magnetic fields respectively, Re the fluid Reynolds number ($r_1^2
\Omega / \nu$, $r_1$ inner radius, $\Omega$ inner sphere rotation
rate, $\nu$ bulk viscosity of the fluid), and Ha the Hartmann number
($B_0 r_1 \sqrt{\sigma / \rho \nu}$, $B_0$ applied field strength,
$\sigma$ electrical conductivity, $\rho$ mass density).

The magnetic field is split into an applied ($\mb{B}_0$) and an
induced ($\mb b$) component, where the applied field is curl free
within the flow domain. The Lorentz force is then given by
\begin{equation}
  \left(\curl \mb B\right) \times \mb B = \left(\curl \mb{b}\right)
  \times \mb{B}_0 + \left(\curl \mb{b}\right) \times
  \mb{b} \label{eqn:LorentzForce},
\end{equation}
\noindent where $\mb b$ is given by the magnetic induction equation in
the (so-called inductionless) limit where diffusion $\left(\nabla^2
\mb b\right)$ exactly balances advection $\left(\curl \left(\mb U
\times \mb B_0\right)\right)$:
\begin{equation}
  0 = \nabla^2 \mb b + \curl \left(\mb U \times \mb{B}_0\right) \label{eqn:Induction}. 
\end{equation}
\noindent The $\left(\left(\curl \mb b\right) \times \mb b\right)$
term in Eqn.~\ref{eqn:LorentzForce} is taken to be
small. \responsenote{2.3} The inductionless limit is valid at low
magnetic Reynolds number
\begin{equation*}
  {\rm Rm} \equiv \frac{\tau_{\rm diff}}{\tau_{\rm eddy}} = \frac{L^2
    / \eta}{L / U_0} \ll 1,
\end{equation*}
\noindent where $\tau_{\rm diff}$ is the magnetic diffusion time,
$\tau_{\rm eddy}$ is the large eddy turn over time, $L$ is the
characteristic scale length, $\eta$ is the magnetic diffusivity, and
$U_0$ is the characteristic velocity. \responsenote{2.4} This implies
that magnetic fields diffuse away on such rapid time scales, relative
to the flow dynamics, that they can only take the shape/value at a
given instant in time that the flow would induce in that instant
alone. Because the field generated in that instant must take its
energy from the flow, the field generation acts as an extra drag on
the development of the flow (akin, if not identical, to visocity).

The flow is simulated using a code, described in
\cite{Hollerbach.IJNMF.2000}, that defines the magnetic and velocity
fields spectrally, in terms of vector spherical harmonics divided into
toroidal and poloidal components. The magnetic boundaries are taken to
be insulating (zero toroidal magnetic field outside the flow, zero
jump in poloidal field at the boundaries); the flow is taken to be
no-slip at the inner and outer boundaries. This paper concerns itself
with only the azimuthal component of the spectral decomposition, and
with interactions between different azimuthal flow modes. The
simulations presented herein were run with spectral resolutions of 60
radial modes, 200 latitudinal modes and 20 longitudinal modes,
consistent with other publications \cite{Travnikov.ActaMech.2011} on
the topic.

The code treats (\ref{eqn:NavierStokes}) pseudospectrally, with the
spectra being expanded out into real space to calculate
(\ref{eqn:Forcing}) and (\ref{eqn:Induction}) and then transformed
back. This is a quite normal method and usually the most efficient way
to go about solving the problem (multiplications are easy in real
space, derivatives are easy in spectral space). If the problem were
treated spectrally, the computer time per time step would increase,
but the flow would evolve indentically to the pseudospectral code. The
analysis presented below takes individual time steps of the
pseudospectral code, and then interprets the dynamics at these time
steps in terms of three-wave coupling of spectra. See
Appendix~\ref{Appendix:TaylorExpansion} for a detailed explanation of
this process. 

The rest of the paper will proceed as
follows. Section~\ref{section:CharacterizingInteractions} provides the
definition of the nonlinear interactions and an introduction to the
nomenclature used to describe them. Sections
\ref{section:ShercliffLayer} and \ref{section:ReturnFlow} below
contain analyses based on networks of nonlinear interactions for the
Shercliff layer instability and the return flow instability
respectively. Section \ref{section:Conclusions} concludes the paper.
\section{Characterizing Interactions}
\label{section:CharacterizingInteractions}

When considering the solution to a nonlinear differential equation one
typically looks for some characterizing value from which a meaningful
interpretation of the evolution can be made. Previous magnetized
spherical-Couette studies \cite{Hollerbach.RSPA.2001,
  Gissinger.PRE.2011} drew their conclusions from the torque on the
outer sphere, in part because a physical experiment would have access
to that measurement directly. Other studies
\cite{Hollerbach.RSPA.2009, Travnikov.ActaMech.2011} used the energies
contained in individual azimuthal modes to demonstrate transitions
between different states as the nondimensional parameters were
varied. Here, we're going to propose the three-wave coupling between
azimuthal modes, complex as it may be to fully consider, as the
relevant characterization. A similar characterization was done in
\cite{Kaplan.PRE.2012} for a kinematic dynamo problem. There the
velocity field catalyzed the interactions of magnetic modes, but was
itself unaffected (as per the definition of kinematic dynamo problem);
here the velocity modes are the catalysts {\em and} the reactants.

\responsenote{1.1} We start by defining an inner product, which
mode energies and energy transfers will be defined with, by the volume
integral of two vector fields dotted together
\begin{equation}
  \innerprod{ \mb{A}}{ \mb{B}} = \!\!\int \limits_1^{1/\eta} \!\!\!
  dr\, r^2 \!\!\!\int \limits_0^\pi \!\!d\theta\, \sin \theta
  \!\!\!\int \limits_0^{2\pi} \!\!d\phi\, \mb{A}(r,\theta,\phi) \cdot
  \mb{B}(r,\theta,\phi) \label{eqn:innerprod}.
\end{equation}
\noindent From this we define the energy contained in each mode

\begin{equation}
  E^{m} = \frac{1}{2} \innerprod{\mb{U}^m}{\mb{U}^m} \label{eqn:energy},
\end{equation}
\noindent 

\noindent with the change in energy in a given mode from some small change 
given by the Taylor expansion:

\begin{equation}
  \delta E^{m} = \innerprod{\mb{U}^m}{\bm{\delta}^m} \label{eqn:deltaE},
\end{equation}
\noindent where $\bm{\delta}^m$ is a small perturbation to the velocity
field of azimuthal mode $m$. 

The individual $\bm{\delta}^m$s to be considered come from the forcing
eqn. (\ref{eqn:Forcing}), which can be broken up into a collection of
interactions between individual $m$ modes represented by the effect of
the coupling on the target mode

\begin{equation}
  \left(a,b \rightarrow c\right) = \left\langle\left(\left(\curl \mb
  U^a \right) \times \mb U^b + \left(\curl \mb U^b \right) \times \mb
  U^a\right), \mb U^c \right\rangle. \label{eqn:coupling}
\end{equation}

\noindent {\em I.e.} the addition or reduction of energy in mode $c$
from the beating of modes $a$ and $b$ defines $\beat{a}{b}{c}$. The
energy dynamics can also be considered in terms of a transfer from one
mode to another, mediated by a third. This is represented by
$\edge{a}{b}{c}$, which represents energy being removed from mode $a$
and deposited in mode $c$ through the interaction with mode $b$. To
use the language of graph theory, $\xrightarrow{b}$ is an edge
connecting two nodes $a$ and $c$. Typically $\edge{c}{a}{b}$ and
$\edge{b}{c}{a}$ are also nonzero. Throughout this paper, interaction
will be used as a general term for both beats and edges.

\responsenote{3.2} This analysis assumes that the change in energy in
a given mode during a given time step is well represented by the
linear sum of individual nonlinear interactions between modes
\begin{align}
 {\gamma}c \equiv \pd{E^c}{t} \approx \frac{E^c(t+\delta
   t)-E^c(t)}{\delta t} \approx \sum \limits_{a,b}
 \frac{\beat{a}{b}{c}}{\delta t} \label{eqn:engsum};
\end{align}

\noindent that the interactions between a given triplet of modes (a,b,
and c) only act to redistribute energy amongst them
\begin{equation}
  \beat{a}{b}{c} + \beat{b}{c}{a} + \beat{c}{a}{b} = 0 \label{eqn:triplets};
\end{equation}
\noindent and that energy is only added to or removed from the
instability through interaction with the axisymmetric background

\begin{equation}
  \pd{\sum \limits_m E^m}{t} = \sum \limits_m \beat{m}{0}{m}, \label{eqn:axisym}
\end{equation}
\noindent where all other interactions only act to redistribute energy
between the various modes. All three assumptions are checked
numerically as the analysis code is run and have heretofor held to
within one percent.

\responsenote{3.1} The consequence of (\ref{eqn:triplets}) is that any
given triplet of beats can be represented entirely by two edges. For a
triplet with $a \neq b \neq c$, one beat of $\beat{a}{b}{c}$,
$\beat{b}{c}{a}$, and $\beat{c}{a}{b}$ will have a larger magnitude
than and opposite sign to the others. For $\beat{a}{b}{c} > 0$, mode
$c$ is acting as an energy sink, and drawing energy (unevenly) from
modes $a$ and $b$. This can be represented by two edges
\begin{align*}
  \edge{a}{b}{c} &= -\beat{b}{c}{a},\quad {\rm and} \\
  \edge{b}{a}{c} &= -\beat{c}{a}{b}, \\
\end{align*}
\noindent where $\edge{m_1}{m_2}{m_3}$ is the energy drawn from $m_1$
and deposited in $m_3$ from the triplet. For $\beat{a}{b}{c}
< 0$, mode $c$ is acting as an energy sink and depositing energy
(unevenly) into modes $a$ and $b$, which is represented as 

\begin{align*}
  \edge{c}{b}{a} &= \beat{b}{c}{a},\quad {\rm and} \\
  \edge{c}{a}{b} &= \beat{c}{a}{b}. \\
\end{align*}

\noindent The transfer from $m_1$ to $m_2$ or {\em vice verse} is
accounted for by the difference between $\edge{m_1}{m_2}{m_3}$ and
$\edge{m_2}{m_1}{m_3}$.


When describing the edges, all diffusive effects are included with the
axisymmetric flow, {\it i.e} as a part of $\edge{a}{0}{a}$. While
viscous diffusion as accounted for in the code communicates energy
between radial modes/nodes, it does not communicate between different
latitudinal or azimuthal modes.  In the discussion to follow, magnetic
effects will also be included as a part of $\edge{a}{0}{a}$. Because
$\mb{B}_0$ is axisymmetric in all simulations presented, both $\curl
\left(\mb{U}^a \times \bm{B}_0\right)$ and $\left(\curl
\mb{b}^a\right) \times \mb{B}_0$ have zero projection onto any mode $m
\neq a$. Similarly, if quadratic effects of the induced magnetic
field (the $\left(\curl{\mb{b}}\right) \times \mb{b}$ term in
Eqn.~\ref{eqn:LorentzForce}) were included, they would be folded into the
$\edge{a}{a}{2a}$ and $\edge{a}{a}{0}$ edges.

\responsenote{2.3} The purpose of these assumptions is to allow the
nonlinear dynamics of the flow to be represented in terms of a network
of interactions. This network formulation is applicable for any system
where there is some global quantity, {\em e.g.} energy or helicity,
for whom the presence in a given mode of a dynamic system is
quantifiable, and for whom the transfer of this quantity between modes
is also quantifiable. Once a list of edges and nodes has been
generated, there are open source tools to visualize the graph. Here we
make use of \texttt{Graphviz}\cite{Gansner2000}.

Sections \ref{section:ShercliffLayer} and \ref{section:ReturnFlow}
below contain analyses based on such networks. In these analyses there
is sometimes reference made to ${\lambda}m$. This represents a guess
of the growth rate of mode $m$ based on
\begin{equation}
  {\lambda}m = e_m E^m, \label{eqn:lingrowth}
\end{equation}
\noindent where $e_m$ is the eigenvalue of the
fastest-growing/slowest-decaying eigenmode of the LNSE analysis. The
guess assumes that the flow found for the given $m$ in the fully-3d
calculation is identical to that eigenmode. This is a good assumption
only if $\edge{m}{0}{m} = {\lambda}m = {\gamma}m$.

As the Shercliff layer instability has the simplest network, that is
where we shall begin.

\section{Saturation of the Shercliff layer instability}
\label{section:ShercliffLayer}

The Shercliff layer is a shear layer that arises in spherical Couette
flows where the magnetic field is strong enough to force the fluid
inside the inner spheres tangent cylinder to corotate with the inner
sphere. The fluid outside the tangent cylinder is in corotation with
the outer sphere ({\it i.e}, at rest in our simulations). The
instability that can arise in this context (see
Fig.~\ref{fig:profiles}c \& f) is akin to a Kelvin-Helmholtz
instability, and was studied fairly extensively in
\cite{Hollerbach.RSPA.2001}.

\begin{figure}
  \centering
  \includegraphics[width=3in]{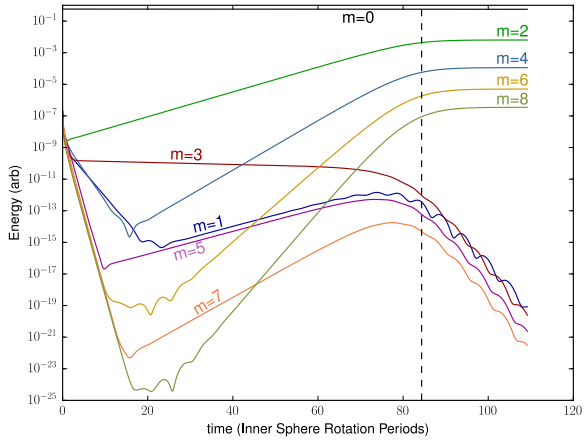}
  \caption{(Color online) \label{fig:Shercliff:Energy} Time series of energies contained
    in each azimuthal mode for a simulation with $\eta$ = 0.5, Re =
    1000, Ha = 70. The vertical line indicates the time slice the
    network diagram in Fig.~\ref{fig:Shercliff:Network} is made from.}
\end{figure}

Fig.~\ref{fig:Shercliff:Energy} shows the energy content of each
azimuthal mode for a run with $\eta=0.5$, Re=1000, Ha=70. This is
where the $m=2$ mode is unstable, but the $m=3$ mode is (just barely)
still stable. The bulk of the energy lives in the $m=2$ azimuthal
mode, which grows exponentially and then begins to asymptote around
t=80. The higher harmonics of $m=2$ grow alongside the first harmonic,
and begin to saturate at the same time. The energies in the odd modes
are all much much smaller than those in the even modes, and as the
even modes asymptote, the odd modes begin to decay roughly
exponentially.

\begin{figure}
  \centering
  \includegraphics[width=3in]{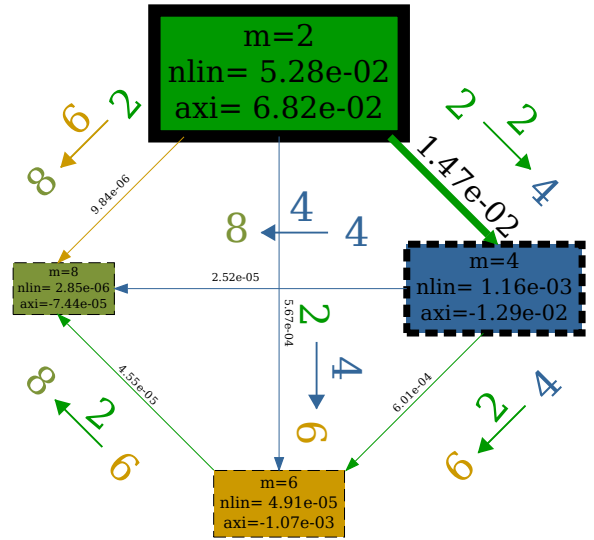}
  \caption{\label{fig:Shercliff:Network} Network of interactions for
    the time indicated in
    Figs.~\ref{fig:Shercliff:Energy},~\ref{fig:Shercliff:M2},~and~\ref{fig:Shercliff:M3}.
    The diagram should be interpreted as follows. The color of each
    arrow indicates $b$ in $\edge{a}{b}{c}$. Here the interactions are
    also written out explicitly along the edges. The size of each box
    scales with the logarithm of the energy in the mode at the given
    time step. The numbers indicated by nlin and axi are,
    respectively, the $\pd{E^m}{t}$ of the mode for the full
    simulation and the influence of the axisymmetric component
    $\edge{m}{0}{m}$.  The strength of the connection $a
    \xrightarrow{b} c$ is written along the edge. The width of each
    edge scales with the logarithm of the connection strength. The
    black border of each node is scaled with the logarithm of the
    magnitude of $\edge{m}{0}{m}$, with dashed lines indicating an
    energy sink and a solid line indicating an energy source. The
    nodes are limited to $a,c \in \{2, 4, 6, 8\}$, the edges are
    limited to $b \in \{2, 4, 6\}]$. }

\end{figure}

\begin{figure}
  \centering
  \includegraphics[width=3in]{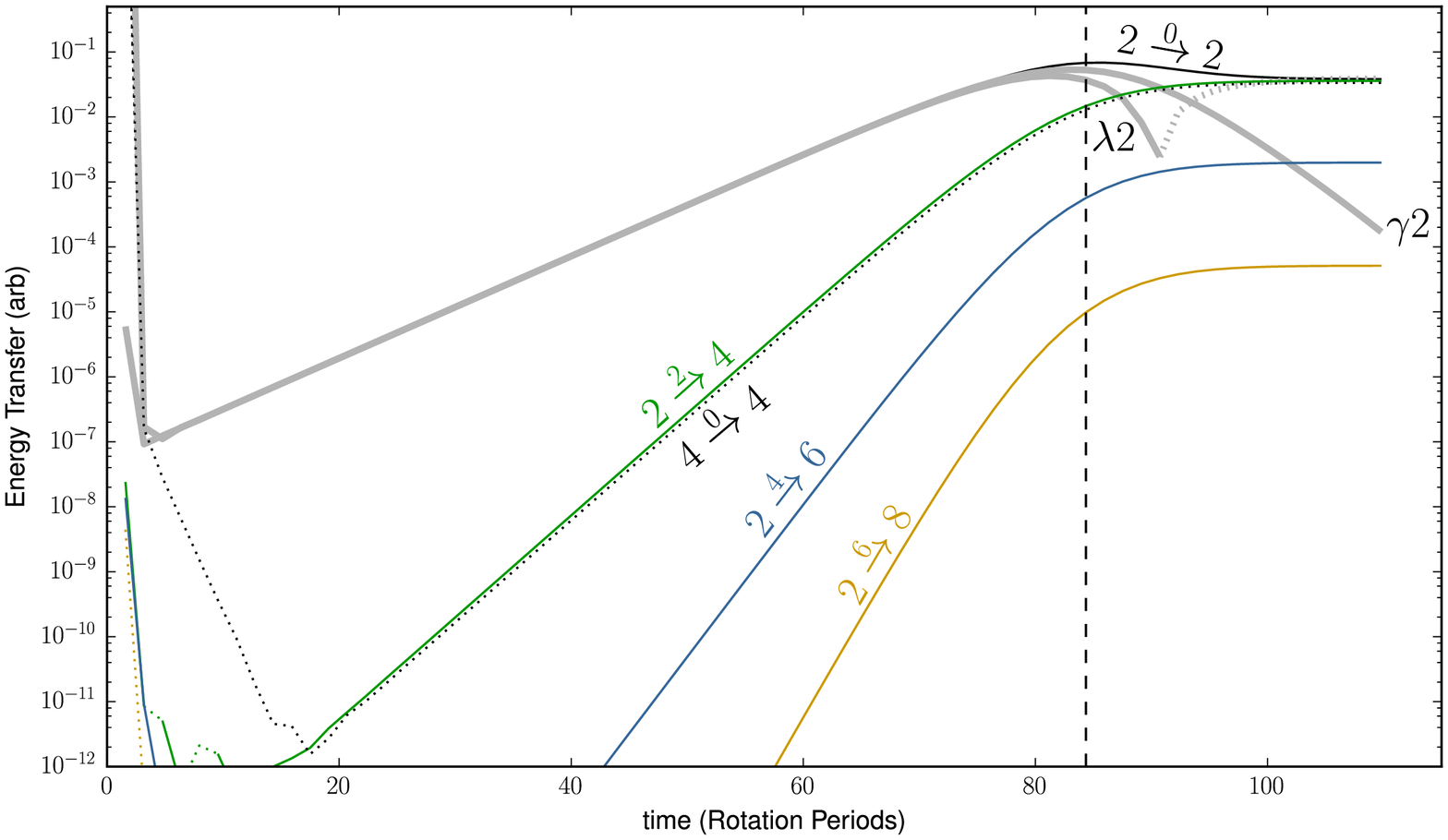}
  \caption{(Color online)\label{fig:Shercliff:M2} Timeseries of a
    subset of the edges from the simulation of
    Fig.~\ref{fig:Shercliff:Energy} on a semilog plot. The dotted
    lines indicate a negative value. The bold lines labeled
    ${\gamma}2$ and ${\lambda}2$ represent the growth rates defined by
    Eqns.~\ref{eqn:engsum} and~\ref{eqn:lingrowth} respectively. The
    interaction $\edge{4}{0}{4}$ represents a decay slightly faster
    than that predicted by ${\lambda}4$ (not pictured). The vertical
    dashed line indicates the timestep
    Fig.~\ref{fig:Shercliff:Network} was made from.}
\end{figure}

Fig.~\ref{fig:Shercliff:Network} shows the network of interactions at
a point during the saturation phase of the instability. This is made
up only of the harmonics of $m=2$. Herein lie examples of most of the
types of edges that will be of interest. \responsenote{2.5} For
example, $\edge{2}{2}{4}$ and $\edge{4}{4}{8}$ represent modes
interacting with themselves nonlinearly and depositing energy into
their second harmonic. Modes $m=6$ and $m=8$ are both acting as sinks
$\left[\sink{2}{6}{4}, \sink{2}{8}{6}\right]$ The vast majority of the
dynamics are contained in the $\edge{2}{0}{2}, \edge{2}{2}{4},$ and
$\edge{4}{0}{4}$ edges. This dominance is demonstrated more clearly in
Fig.~\ref{fig:Shercliff:M2}, where the growth of $m=2$ is
indistinguishable from $\edge{2}{0}{2}$ until $t\approx80$. At this
point $\edge{2}{2}{4}$ is on the same order of magnitude as
$\edge{2}{0}{2}$, and from there on out the two edges asymptote
towards each other. The energy being deposited in $m=4$ is almost
completely dissipated by the background flow. The next largest edge,
($(\edge{4}{2}{6}$, not shown) has an impact an order of magnitude
weaker than $\edge{4}{0}{4}$.

\begin{figure}
  \centering
  \includegraphics[width=3in]{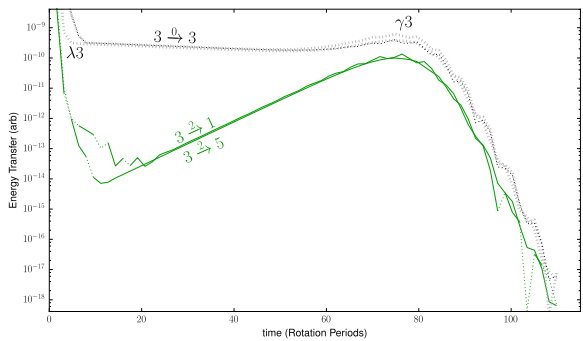}
  \caption{(Color online) \label{fig:Shercliff:M3} Timeseries of a subset of the
    edges from the simulation of Fig.~\ref{fig:Shercliff:Energy} on a
    semilog plot. The dotted lines indicate a negative value. The bold lines labeled
    ${\gamma}3$ and ${\lambda}3$ represent the growth rates defined by
    Eqns.~\ref{eqn:engsum} and~\ref{eqn:lingrowth} respectively. The
    vertical dashed line indicates the timestep
    Fig.~\ref{fig:Shercliff:Network} was made from.}
\end{figure}

The odd modes are even simpler. Fig.~\ref{fig:Shercliff:M3} shows a
time series of the edges relevant for $m=3$, and there are few. Until
$t\approx60$ the decay rate is indistinguishable from
$\edge{3}{0}{3}$, at which point $\left(1 \xleftarrow{2} 3
\xrightarrow{2} {5}\right)$ is large enough to notice on the log
scale. However, once the $m=2$ harmonics start saturating the
$m_{odd}$ modes crash, and the decay rate returns to being a near
match of $\edge{3}{0}{3}$.

\section{Saturation of the return flow instability}
\label{section:ReturnFlow}

At lower Ha, the equatorial jet is no longer suppressed, but neither
does it reach the edge of the sphere. Instead it returns somewhere in
between $r_1$ and $r_2$ with a stagnation point on the equator. The
return flow instability arises in this stagnation region (see
Fig.~\ref{fig:profiles}b \& e).

\begin{figure}
  \centering
  \includegraphics[width=3in]{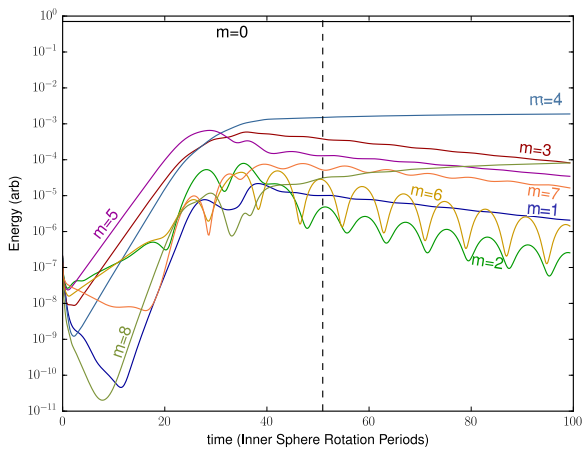}
  \caption{(Color online) \label{fig:Return:Energies} Time series of energies
    contained in each azimuthal mode for a simulation with $\eta$ =
    0.5, Re = 1000, Ha = 30. The vertical line indicates the time
    slice the network diagram in Fig.~\ref{fig:fullnetwork} is made
    from. }
\end{figure}

The two dynamics that the network characterization seeks to describe
are the saturation of the dominant mode, and the suppression of the
subdominant modes that are still linearly unstable. In both cases it
can be shown that the network of interactions transfers energy from
the unstable modes (wherein it is created) to stable modes (wherein it
is destroyed). Fig.~\ref{fig:Return:Energies} shows the evolution of
the flow from an initial state (found by evolving an axisymmetric flow
with Re=1000, Ha=30, $\eta$=0.5 to steady state), seeded with random
nonaxisymmetric noise, to what is taken to be saturation. Up until
t$\approx$15, $m \in [2,6]$ seem to all be growing
exponentially. Until about t $\approx$ 25, $m \in [3,5]$ continue to
grow roughly exponentially. From t $\approx$ 25 on several changes can
be observed. First $m=5$ rolls over and begins to decay, then $m=3$
rolls over and begins to decay as $m=4$ begins to saturate.

\begin{figure*}
  \centering
  \includegraphics[width=6in]{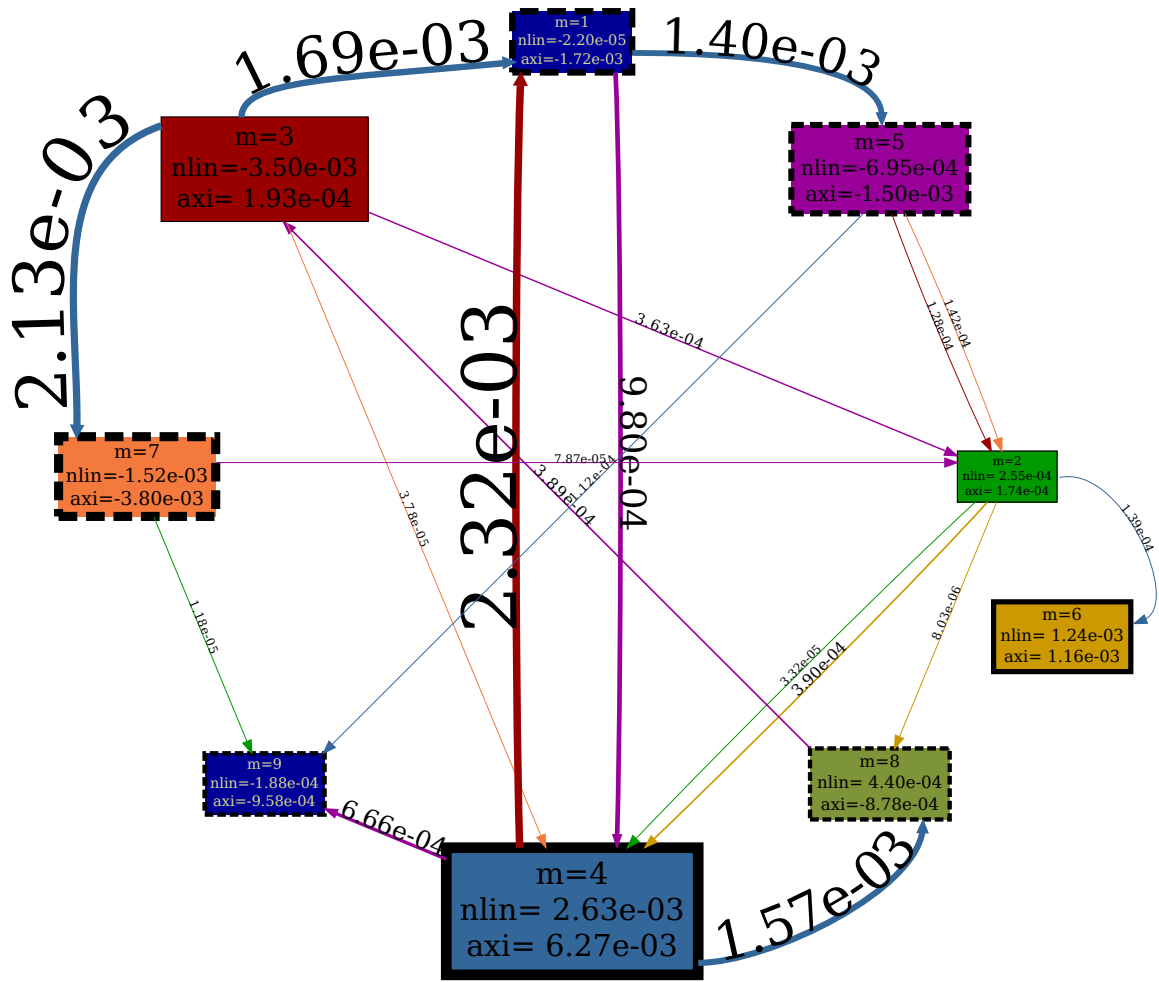}
  \caption{\label{fig:fullnetwork} Full network of interactions from
    the timestep indicated in Fig.~\ref{fig:Return:Energies}. The
    diagram should be interpreted as in
    Fig.~\ref{fig:Shercliff:Network}. The nodes are limited to $a,c
    \in [1,9]$, the edges are limited to $b \in [2, 7]$. A tabular
    form of the information is in
    Appendix~\ref{Appendix:InteractionTable}; Table~\ref{table:nodes}
    contains the information contained in the nodes;
    Table~\ref{table:edges} contains a list of edge strengths.}
\end{figure*}

This is where the network formulation comes into
play. Fig.~\ref{fig:fullnetwork} shows the network of interactions at
a single point in time. Several dynamics are visible here. The $m=4$
mode is sourcing energy and depositing much of it in it's
2\textsuperscript{nd} harmonic via $\edge{4}{4}{8}$. The $m=3$ mode
sources energy as well, but it is a net looser of energy as more is
being sent to energy dissipating modes via $\left(7 \xleftarrow{4} 3
\xrightarrow{4} 1\right)$. The $m=1$ mode dissipates energy but is
likely more significant as a path for energy to move between $m=3, 4,
$ and $5$.

\begin{figure}
  \centering
  \includegraphics[width=3in]{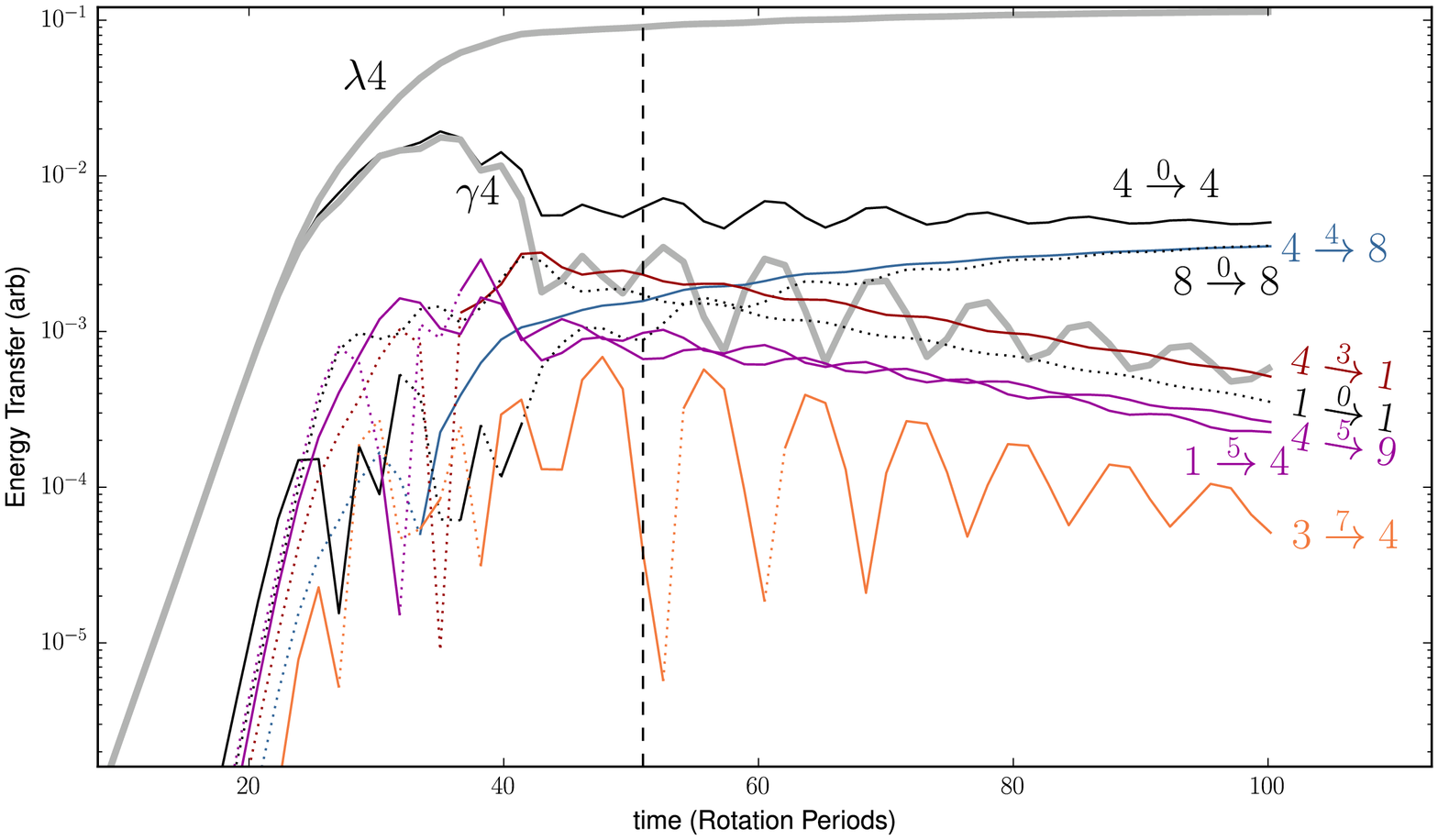}
  \caption{(Color online)\label{fig:Return:m4_saturation} Timeseries
    of a subset of the edges from the simulation of
    Fig.~\ref{fig:Return:Energies} on a semilog plot. The dotted lines
    indicate a negative value. The bold lines labeled ${\gamma}4$ and
    ${\lambda}4$ represent the growth rates defined by
    Eqns.~\ref{eqn:engsum} and~\ref{eqn:lingrowth} respectively. The
    vertical line indicates the time slice the network diagram in
    Fig.~\ref{fig:fullnetwork} is made from.}
\end{figure}

From the diagram we choose interesting edges to track over
time. Fig.~\ref{fig:Return:m4_saturation} shows the dominant
edges which transfer energy to or from the $m=4$ mode, with the
addition of the total growth rate, and the $\edge{8}{0}{8}$ and
$\edge{1}{0}{1}$ edges. Up until $t \approx 25$, there is
exponential growth which is almost entirely identical to the
$\edge{4}{0}{4}$ term. Until $t \approx 40$, the growth is still
almost entirely identical to the $\edge{4}{0}{4}$ term, but this
edge has begun to roll over and asymptote. There are 4
edges between $t \approx 40$ and the end of the simulation that
account for the vast majority of the dynamics of the $m=4$ mode:
$\edge{4}{3}{1}$ initially draws the largest part of the energy from
$m=4$; $\edge{4}{4}{8}$ dominates at long
times; $\edge{4}{5}{9}$ and $\edge{1}{5}{4}$ are roughly equal,
indicating that they are better considered as a single action $\left(1
\xrightarrow{5} 4 \xrightarrow{5} 9\right)$ which does not matter much
in the energy dynamics of $m=4$ itself. All but a small, and diminishing,
component of the energy transfered to $m=8$ is removed by
$\edge{8}{0}{8}$. There is, on the other hand, a rather stable
relationship between the amount of energy transfered into $m=1$ by
$\edge{4}{3}{1}$, the amount removed by $\edge{1}{0}{1}$ and the total
growth rate $\pd{E^{m=4}}{t}$

\begin{figure}
  \centering
  \includegraphics[width=3in]{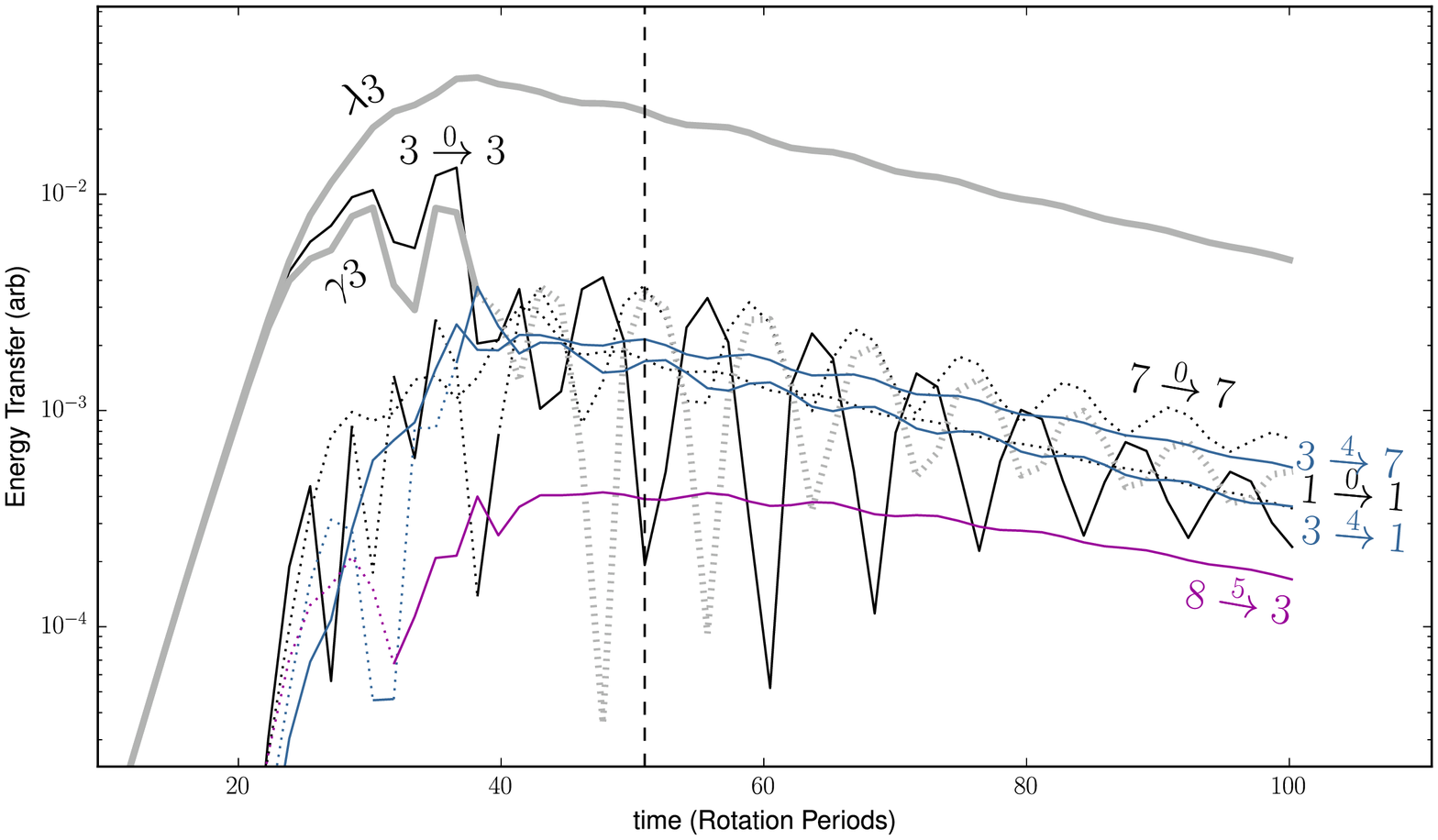}
  \caption{(Color online) \label{fig:Return:m3_decay} Timeseries of a
    subset of the threewave couplings from the simulation of
    Fig.~\ref{fig:Return:Energies} on a semilog plot. The dotted lines
    indicate a negative value. The bold lines labeled ${\gamma}3$ and
    ${\lambda}3$ represent the growth rates defined by
    Eqns.~\ref{eqn:engsum} and~\ref{eqn:lingrowth} respectively. The
    vertical line indicates the time slice the network diagram in
    Fig.~\ref{fig:fullnetwork} is made from.}
\end{figure}

The rest of the story is contained in
Fig.~\ref{fig:Return:m3_decay}. Like $m=4$, $m=3$ grows exponentially
until $t \approx 25$ from $\edge{3}{0}{3}$. However, a gap opens up
between the total growth rate and the energy drawn from the
axisymmetric flow here, and from $t \approx 40$ onward there is a net
loss of energy from the $m=3$ mode, despite the fact that the mean
flow is a constant source of energy.

The majority of the energy flow out of $m=3$ is described by
$\edge{3}{4}{7}$ and $\edge{3}{4}{1}$. The $m=7$ mode is super stable,
and loses more energy to the background flow than is deposited by
$\edge{3}{4}{7}$. The $m=1$ mode is also stable, but its energy
dissipation tends to match the energy deposited by $\edge{3}{4}{1}$
almost exactly.
\section{Conclusions}
\label{section:Conclusions}

The saturation of the Shercliff layer ($\eta$ 0.5, Re 1000, Ha 70) and
return flow ($\eta$ 0.5, Re 1000, Ha 30) instabilities are
characterized by the three-wave coupling between azimuthal modes
($m$). In both cases energy is generated by $\edge{m}{0}{m}$,
dissipated by $\edge{2m}{0}{2m}$, and transferred between the two by
$\edge{m}{m}{2m}$. Furthermore, in the case of the return flow
instability, the dominant mode suppresses other unstable modes by
facilitating a transfer of energy into higher order modes which then
dissipate the energy.  This suppression is a possible candidate for the
hysteresis cyles of \cite{Hollerbach.RSPA.2009}.

\responsenote{2.7} The network diagram is instrumental in this form of
analysis. For a simulation with 20 azumuthal modes, there are 400
possible interactions satisfying $c = \left|a \pm b\right|$. The
diagram provides a snapshot of interactions meeting certain criteria
(nodes or edges as members of a chosen set, displayed edges accounting
for 90\% of the total energy flow). Interactions can be picked from
this snapshot and tracked throughout the simulation to see how they
evolve and how they relate to the final saturated state.

This formulation can be extended, at a price. Here we defined the
nodes only by azimuthal order $m$; in \cite{Kaplan.PRE.2012} the nodes
were further divided into degree ($l$), phase (sin and cos) and
toroidal/poloidal character. This was sensible there because the
number of distinct edges was limited to the (four) harmonics of the
defined flow, and because individual edges or series of edges could be
connected to the $\alpha$ and $\Omega$ effects of dynamo theory. The
case of magnetized inductionless spherical Couette flow would most
likely not benefit from the full decomposition. However, it may still
be meaningful to distinguish between equatorially symmetric ($l \in
[m,m+2,\ldots,l_{max}]$) and antisymmetric ($l \in
[m+1,m+3,\ldots,l_{max}-1]$) modes, as these classes of flow modes are
excited or suppressed in different regions of the (Re, Ha) phase
space.


The work presented here and that presented in \cite{Kaplan.PRE.2012}
only cover the cases where energy transfer is facilitated by velocity
modes (here between the velocity modes themselves, in
\cite{Kaplan.PRE.2012} between magnetic modes). \responsenote{2.3}
This network formulation is applicable for any system where there is
some global quantity, {\em e.g.} energy or helicity, for whom the
presence in a given mode of a dynamic system is quantifiable, and for
whom the transfer of this quantity between modes is also
quantifiable. As a further example, one could consider a saturating
dynamo. A typical simulation, such as those in \cite{Reuter.NJP.2009},
will show anticorrelations between the energies in the velocity and
magnetic fields, which implies that there is energy being transferred
between them. The primary decisions are how the modes are defined and
how the edges are calculated.

\begin{acknowledgments}
  The author would like to thank Rainer Hollerbach for providing the
  source code from \cite{Hollerbach.IJNMF.2000}, Rainer Hollerbach and
  Andre Giesecke are also thanked for acting as 'round 0' reviewers
  for the manuscript. This work was supported as part of the DRESDYN
  project\cite{DRESDYN} under Frank Stefani at the Helmholtz-Zentrum
  Dresden-Rossendorf.  This work is supported by the Deutsche
  Forschungsgemeinschaft under grant STE 991/1-2.
\end{acknowledgments}
\appendix
\section{Taylor Expansion of Nonlinear Interactions}
\label{Appendix:TaylorExpansion}

Hollerbach \cite{Hollerbach.IJNMF.2000} describes the time evolution
of the velocity field in terms of a modified 2\textsuperscript{nd} order Runga-Kutta method, with 

\begin{align}
  \mb X \mathbf{\tilde v}(t+\delta t) &= \mb Y \mb v(t) + \delta t \mathbf{DV}, \quad {\rm and} \label{eqn:predictor}\\ 
  \mb X \mb v(t+\delta t) &=  \mb Y \mb v(t) + \frac{\delta t}{2}  \left(\mathbf{DV}^\prime +\mathbf{DV} \right), \label{eqn:corrector}
\end{align}
\noindent with $\mb v$ comprising both the toroidal and poloidal
modes ($\mb e$ and $\mb f$ in
\cite{Hollerbach.IJNMF.2000}), $\mb X$ and $\mb Y$ operators
that only connect $k$ terms in the spectra with the same $l$ and $m$,
and $\mathbf{DV}$ and $\mathbf{DV}^\prime$ representing the forcing on
a given $k,l,m$ spectrum. For the purposes of the Taylor expansion we
are only going to deal with the predictor term. The forcing is given
by
\begin{align}
  \mb F^{a,b}(r,\theta,\phi) = {\rm Re} \left(\curl \mb
  U^a(r,\theta,\phi)\right) & \times \mb U^b(r,\theta,\phi) \nonumber
  \\ + {\rm Re} \left(\curl \mb U^b(r,\theta,\phi)\right) & \times \mb
  U^a(r,\theta,\phi). \label{eqn:forcing2}
\end{align}
\noindent There are three transformations to get from the spectral
representation the flow is stored in to the spatial representation the
force is calculated in:
\begin{align}
  \mb{U}^m_l(r) &= \sum \limits_k T_{k,l}(r) \mb{v}^m_{k,l}  \label{eqn:klm2rlm}\\
  \mb{U}^m(r,\theta) &= \sum \limits_l P^m_l(\theta) \mb{U}^m_l(r) \label{eqn:rlm2rtm}\\
  \mb{U}(r,\theta,\phi) &= \mathcal{F}^{-1}\left\{ \mb{U}^m(r,\theta) \right\} \label{eqn:rtm2rtp}.
\end{align}
\noindent where $\mathcal{F}^{-1}$ is an inverse Fourier transform,
$P^m_{l}$ is an expansion in associated Legendre polynomials, and
$T_{k,l}$ is an expansion in Chebyshev polynomial that may be slightly
modified to calculated the curl of the spectrum. After the forcing is
calculated in real space it is reverted to the spectral representation
through another 3 transformations:
\begin{align}
  \mb{F}^m(r,\theta) &= \mathcal{F}\left\{ \mb{F}(r,\theta, \phi) \right\} \label{eqn:rtp2rtm} \\
  \mb{F}^m_l(r) &= \sum \limits_l \mathcal{P}^m_l(\theta) \mb F^{m}(r,\theta) \label{eqn:rtm2rlm}{\rm and} \\ 
  \mb{DV}^m_{k,l} &= \sum \limits_k \mathcal{T}_{k,l}(r) \mb {F}^m_l(r)  \label{eqn:rlm2klm},\\
\end{align}

\noindent where $\mathcal F$ is a fourier transform, and
$\mathcal{P}^m_l$ and $\mathcal T_{k,l}$ transform the spatial
function into Chebyshev and Legendre spectra with some curls of $\mb
F$ included. The only place where there is communication between m
modes is between \ref{eqn:rtm2rtp} and \ref{eqn:rtp2rtm}, which allows
us to treat the predictor step (Eqn.~\ref{eqn:predictor}) as

\begin{equation}
  \mb X \tilde v^m(t+\delta t) = \mb Y v^m(t) + \delta t \sum \limits_{a,b} \mathbf{DV}^{a,b,m}
\end{equation}
where $\mathbf{DV}^{a,b,m}$ is the forcing from (\ref{eqn:forcing2}),
projected onto $m$. The three-wave coupling defined in
Eqn.~\ref{eqn:coupling} is given by
\begin{equation*}
  \beat{a}{b}{m} = \delta t \mb X^{-1} \mathbf{DV}^{a,b,m},
\end{equation*}
\noindent which is nonzero only for $m = \left|a \pm b\right|$. 
\section{Network of interactions in table form}
\label{Appendix:InteractionTable}
The networks of
Figs.~\ref{fig:Shercliff:Network}~and~\ref{fig:fullnetwork} are
difficult to read. They are, however, a good graphical snapshot of
where energy is moving in a nonlinear process where there is no
obvious hierarchy of interactions. Table~\ref{table:nodes} could be
sorted by mode index, as it is, or by any of the entries in the table
and still be easily interpreted; there is no hierarchy of keys in
Table~\ref{table:edges} that reveals multi-step interactions $\left(a
\xrightarrow{b} c \xrightarrow{d} e\right)$ as completely as
Fig.~\ref{fig:fullnetwork}.

\begin{table}
  \begin{tabular}{|l|l|l|l|} 
    \hline mode & energy & nonlinear & axisymmetric \\ \hline 
    1 &  1.01e-05 & -2.20e-05 & -1.72e-03\\ \hline 
    2 &  2.70e-06 &  2.55e-04 &  1.74e-04\\ \hline 
    3 &  4.09e-04 & -3.50e-03 &  1.93e-04\\ \hline 
    4 &  1.47e-03 &  2.63e-03 &  6.27e-03\\ \hline 
    5 &  1.31e-04 & -6.95e-04 & -1.50e-03\\ \hline 
    6 &  2.42e-05 &  1.24e-03 &  1.16e-03\\ \hline 
    7 &  6.34e-05 & -1.52e-03 & -3.80e-03\\ \hline 
    8 &  2.70e-05 &  4.40e-04 & -8.78e-04\\ \hline 
    9 &  1.02e-05 & -1.88e-04 & -9.58e-04\\ \hline 
  \end{tabular} 
  \caption{Tabular form of network diagram of
    Fig.~\ref{fig:fullnetwork}. List of nodes and their associated
    energies, nonlinear growth rates, and the action of the
    axisymmetric flow. The entries are sorted by azimuthal mode
    number.} \label{table:nodes}
\end{table}

\begin{table}
  \begin{tabular}{|l|l||l|l||l|l|} 
    \hline edge & strength & edge & strength & edge & strength  \\ \hline 
    $ 1 \xrightarrow{ 4}  5$ &  1.40e-03 & $ 3 \xrightarrow{ 4}  1$ &  1.69e-03 & $ 6 \xrightarrow{ 3}  3$ &  1.15e-04 \\ \hline 
    $ 1 \xrightarrow{ 5}  4$ &  9.80e-04 & $ 3 \xrightarrow{ 4}  7$ &  2.13e-03 & $ 6 \xrightarrow{ 7}  1$ &  5.23e-05 \\ \hline 
    $ 1 \xrightarrow{ 5}  6$ &  1.70e-04 & $ 3 \xrightarrow{ 5}  2$ &  3.63e-04 & $ 7 \xrightarrow{ 2}  9$ &  1.18e-05 \\ \hline 
    $ 1 \xrightarrow{ 6}  5$ &  1.24e-04 & $ 3 \xrightarrow{ 7}  4$ &  3.78e-05 & $ 7 \xrightarrow{ 5}  2$ &  7.87e-05 \\ \hline 
    $ 2 \xrightarrow{ 2}  4$ &  3.32e-05 & $ 4 \xrightarrow{ 3}  1$ &  2.32e-03 & $ 7 \xrightarrow{ 6}  1$ &  1.12e-04 \\ \hline 
    $ 2 \xrightarrow{ 3}  1$ &  3.20e-05 & $ 4 \xrightarrow{ 4}  8$ &  1.57e-03 & $ 8 \xrightarrow{ 5}  3$ &  3.89e-04 \\ \hline 
    $ 2 \xrightarrow{ 4}  6$ &  1.39e-04 & $ 4 \xrightarrow{ 5}  9$ &  6.66e-04 & $ 8 \xrightarrow{ 7}  1$ &  6.59e-05 \\ \hline 
    $ 2 \xrightarrow{ 6}  4$ &  3.90e-04 & $ 5 \xrightarrow{ 3}  2$ &  1.28e-04 & $ 9 \xrightarrow{ 6}  3$ &  4.31e-05 \\ \hline 
    $ 2 \xrightarrow{ 6}  8$ &  8.03e-06 & $ 5 \xrightarrow{ 4}  9$ &  1.12e-04 & &\\ \hline 
    $ 3 \xrightarrow{ 2}  1$ &  7.79e-05 & $ 5 \xrightarrow{ 7}  2$ &  1.42e-04 & &\\ \hline 
  \end{tabular} 
  \caption{Tabular form of network diagram of
    Fig.~\ref{fig:fullnetwork}. List of edges and their strengths. The
    entries are sorted by source, edge and target.} \label{table:edges}
\end{table}


\end{document}